\documentclass[journal]{IEEEtran}
\usepackage{simplemargins}
\def\ps@headings{%
\def\@oddhead{\mbox{}\scriptsize\rightmark \hfil \thepage}%
\def\@evenhead{\scriptsize\thepage \hfil \leftmark\mbox{}}%
\def\@oddfoot{}%
\def\@evenfoot{}}
\makeatother 
\usepackage{graphicx}
\usepackage{indentfirst}
\usepackage{epsfig}
\usepackage{amsfonts}
\usepackage{subfigure}
\usepackage{url}
\usepackage{authblk}
\usepackage{algorithmic}
\usepackage{algorithm}
\usepackage[cmex10]{amsmath}
\usepackage{amsthm}
\newtheorem{mydef}{Proposition}

\title{Incentive Mechanisms for Hierarchical Spectrum Markets}

\author[1]{George Iosifidis}
\author[2]{Anil Chorppath}
\author[3]{Tansu Alpcan}
\author[1]{Iordanis Koutsopoulos}
\affil[1]{Dep. of Computer and Comm. Eng., Univ. of Thessaly, Greece, } %
\affil[2]{Technical University of Munich, Germany, } %
\affil[3]{Dep. of Electrical and Electronic Eng., University of
Melbourne, Australia}

\begin{document}
\maketitle \setcounter{page}{1} \thispagestyle{empty}
\newtheorem{property}{Property}
\newcommand{\be}{\begin{itemize}} \newcommand{\ee}{\end{itemize}}
\newcommand{\tb}{\textbf} \newcommand{\ttt}{\texttt}
\newcommand{\tit}{\textit} \newcommand{\uline}{\underline}
\newtheorem{proposition}{Proposition}
\newtheorem{theorem}{Theorem} \newtheorem{lemma}{Lemma}
\newtheorem{fact}{Fact}

\begin{abstract}
In this paper, we study spectrum allocation mechanisms in
hierarchical multi-layer markets which are expected to proliferate
in the near future based on the current spectrum policy reform
proposals. We consider a setting where a state agency sells spectrum
channels to Primary Operators (POs) who subsequently resell them to
Secondary Operators (SOs) through auctions. We show that these
hierarchical markets do not result in a socially efficient spectrum
allocation which is aimed by the agency, due to lack of coordination
among the entities in different layers and the inherently selfish
revenue-maximizing strategy of POs. In order to reconcile these
opposing objectives, we propose an incentive mechanism which aligns
the strategy and the actions of the POs with the objective of the
agency, and thus leads to system performance improvement in terms of
social welfare. This pricing-based scheme constitutes a method for
hierarchical market regulation. A basic component of the proposed
incentive mechanism is a novel auction scheme which enables POs to
allocate their spectrum by balancing their derived revenue and the
welfare of the SOs.
\end{abstract}

\section{Introduction}\label{section:1}

\subsection{Background and Motivation}

Nowadays, it is common belief that the current coarse and static
spectrum management policy creates a spectrum shortage. While this
resource is expensive and scarce, significant amount of the reserved
spectrum remains idle and unexploited by legitimate owners. A
prominent proposed solution is the reform of the spectrum allocation
policy and the deployment of dynamic spectrum (DS) markets,
\cite{zhao}. Spectrum should be allocated in a finer spatio-temporal
scale to the interested buyers, the so-called primary operators
(POs), \cite{peha2} and more importantly, the POs should be able to
lease their idle spectrum to secondary operators (SOs), \cite{peha},
who serve fewer users in smaller areas. This method is expected to
increase spectrum utilization and already several related business
models exist in the market, \cite{spectrumbridge}. However, the
market-based solution is not a panacea and should be carefully
applied.

These schemes will give rise to hierarchical spectrum markets where
the spectrum will be allocated in two stages, i.e. from a state
agency to the POs, and from each PO to the SOs. The objective of the
agency, which we call hereafter \emph{controller} (CO), is to
allocate the spectrum efficiently, i.e. so as to maximize the
aggregate social welfare from its use. However, this objective
cannot be achieved by these markets because of the following two
reasons: (i) the \textbf{coordination problem}, and the (ii)
\textbf{objectives misalignment problem}. The first problem emerges
when the CO assigns the spectrum to the POs without considering the
needs of the SOs (secondary demand). The second problem arises due
to the inherently selfish behavior of POs who resell their spectrum
in order to maximize their revenue. Clearly, this strategy
contradicts the goal of the controller.

\begin{figure}[t]
\begin{center}
\epsfig{figure=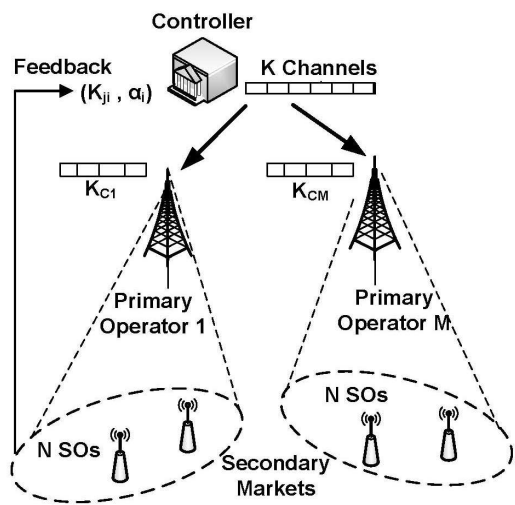, width=7cm,height=7cm}
\end{center}
\caption{The system consists of one Controller which has at his
disposal $K$ identical channels. There exist $M$ primary operators
which ask for spectrum. Each PO $j$ acquires $K_{cj}$ channels and
uses $K_{j0}$ of them to satisfy the needs of his own users and
resells $K_{ji}$ channels to each SO $i$ in the underneath secondary
market. There are $N$ SOs in the monopoly market under each PO which
provide feedback to the CO for their needs and the decisions of the
POs.} \label{fig:ssa1}
\end{figure}

In this paper we study the spectrum allocation in these hierarchical
markets and propose an \emph{incentive mechanism} that enhances
their performance by addressing the above two issues. The mechanism
is deployed by the controller who acts as \emph{regulator} and
incentivizes the POs to redistribute their spectrum in a socially
aware fashion. We consider a basic setting depicted in Figure
\ref{fig:ssa1}, where each PO is a monopolist and has a certain
clientele of SOs. Monopolies are expected to arise very often in
these markets because the POs obtain the exclusive spectrum use
rights for certain areas or because they collude and act effectively
as one single seller. First, we analyze the performance of the
unregulated hierarchical market, i.e. when there is no incentive
mechanism, and we show that it results in an undesirable
equilibrium. The spectrum allocation from the CO to the POs and from
the POs to the SOs is accomplished through auction-based mechanisms
since there is lack of information about the spectrum demand.
Namely, the CO uses an efficient auction such as the VCG auction,
\cite{krishna}, while the POs employ an optimal auction,
\cite{myerson}, which maximizes the expected revenue of the seller
but induces efficiency loss, \cite{hajek}, \cite{maskin}.

Accordingly, we propose a pricing based incentive mechanism where
the CO charges each PO in proportion to the inefficiency that is
caused by his spectrum redistribution decisions. This way, the POs
are induced to allocate their spectrum using a new auction scheme
which produces less revenue for them but more welfare for the SOs.
This is a \emph{novel multi-item auction mechanism} where the
objective of the auctioneer is a linear combination of his revenue
and the valuations of the bidders. The balance between the objective
of the POs and the SOs is tuned by a scalar parameter which is
determined by the CO and reflects his regulation policy. Finally, we
apply our mechanism to dynamic spectrum markets where the CO-POs and
the PO-SOs interactions are realized in different time scale.
Although in this case the coordination problem is inherently
unsolvable, the proposed scheme still improves the performance of
the market by aligning the decisions of the POs with the objective
of the CO.



\subsection{Related Work and Contribution}

Optimal auctions were introduced by Myerson, \cite{myerson} for
single item allocation and extended later for multiple items,
\cite{branco}, or divisible resource, \cite{maskin}. They maximize
the expected revenue of the seller but are inefficient,
\cite{krishna}, \cite{hajek}. The interaction of primary and
secondary operators is usually modeled as a monopoly market. For
example, in \cite{jia} the authors consider a setting where each
primary license holder sells his idle spectrum channels to a set of
secondary users and show that the optimal auction yields higher
profit but results in inefficient allocation. A similar monopolistic
setting is considered in \cite{QZhang} and \cite{zongpeng}. In
\cite{la}, a multiple-item optimal auction is used by a primary
service provider to allocate his channels to a set of secondary
service providers while satisfying at the same time his own needs.
It can be argued that even in oligopoly spectrum markets is highly
probable that the POs - SOs interaction will result in spectrum
allocation that is not efficient from the perspective of the
controller, \cite{courcoubetis}. All these works analyze exclusively
the primary - secondary operators interactions without taking into
account the hierarchical structure of the spectrum markets.

This hierarchical aspect is studied in \cite{niyato} where the
authors consider a multi-level spectrum market and present a
mechanism to match the demand and the spectrum supply of the
interrelated spectrum markets in the different layers. Similar
models have been considered in \cite{huang} and \cite{sarkar} where
the buyers demand is considered known. Coordination problems have
been also studied for bandwidth allocation in wireline networks,
\cite{bitsaki}. However, in these studies there is no misalignment
among the objectives of the various entities (operators, users, etc)
since they all maximize the revenue or the efficiency of the
allocation. \emph{On the contrary, in the setting we study the
entities have conflicting interests and the incentive mechanism we
present achieves their alignment}. Our work is inspired by the
sponsored search (keyword) auction mechanisms, \cite{nisan}, which
assign the search engines advertising slots by taking into account
the feedback from the \emph{clickers}. Similar concepts can be used
for the allocation of spectrum as we suggested in
\cite{iosifidis-commag}. Here, we take a further step towards this
direction by giving a detailed methodology.

In summary, the contributions of this paper are the following: $(1)$
we analyze the hierarchical spectrum allocation and show that it is
inefficient, $(2)$ we present an incentive mechanism that motivates
the POs to increase the efficiency of their spectrum redistribution,
$(3)$ we introduce the $\beta$-optimal auction which achieves a
balance between the revenue of the seller (optimality) and the
welfare of the buyers (efficiency). This is a new mechanism that can
be used also for the allocation of similar communication resources
(bandwidth, transmission power, etc), $(4)$ we apply our mechanism
to dynamic markets where the CO-POs and the PO-SOs interactions are
realized in different time scales and we show that it improves their
efficiency. To our knowledge, this is the first work that
analytically studies the efficiency of the hierarchical spectrum
markets and introduces a mechanism for their performance
improvement.

The rest of the paper is organized as follows. In Section
\ref{section:systemmodel} we introduce the system model and in
Section \ref{section:unregulated} we analyze the hierarchical
spectrum allocation without the intervention of the controller. This
analysis helps us to describe the incentive mechanism and assess its
efficacy in Section \ref{section:regulatedspectrum}. Finally in
Section \ref{section:discussionanddynamicmarkets} we apply our
mechanism to more dynamic spectrum markets. We present our numerical
study in section \ref{section:simulations} and conclude in Section
\ref{section:conclusions}.

\section{System Model} \label{section:systemmodel}
We consider a three-layer hierarchical spectrum market with one
\emph{controller} (CO) on top of the hierarchy, a set
$\mathcal{M}=\{1,2,\ldots,M\}$ of \emph{primary operators} (POs) in
the second layer and a set $\mathcal{N}=\{1,2,\ldots,N\}$ of
\emph{secondary operators} (SOs) that lie in the third layer under
each PO, as it is shown in Figure \ref{fig:ssa1}. There exists a set
$\mathcal{K}=\{1,2,\ldots,K\}$ of identical spectrum channels
managed initially by the CO. The controller allocates the channels
to the $M$ primary operators and accordingly each PO redistributes
the channels he acquired among himself and the $N$ SOs that lie in
his secondary market. The objective of the POs is to incur maximum
revenue from reselling the spectrum while satisfying their own
needs.

The perceived utility of each operator, PO or SO, for acquiring a
channel is represented by a scalar value. Following the law of
diminishing marginal returns, \cite{courcoubetis}, we consider that
each additional channel has smaller value/benefit for the operator.
Different operators may have different spectrum needs and hence
different channel valuations. For example, an operator with many
clients will have very high channel valuations. Also, the POs are
expected to have in general higher valuations than the SOs since
they serve more users. We summarize these different characteristics
of the operators with a real-valued parameter which we call the
\emph{type} of the operator, \cite{krishna}, \cite{la}. Notice that,
our system model is general and satisfies the basic assumptions and
requirements of many different settings, \cite{QZhang},
\cite{niyato}, \cite{huang}, \cite{bitsaki}.

\textbf{Secondary Operators:} In detail, the perceived utility of
the $i^{th}$ SO for the $k^{th}$ channel is $U_{k}(\alpha_{i})\,
\in\, \mathcal{R}^{+}$ which is assumed to be positive,
monotonically increasing and differentiable function of parameter
$\alpha_i$. This is the \emph{type} of the SO and represents his
spectrum needs. The types of the SOs in every secondary market
$\mathbf{\alpha}=(\alpha_i:\,i\in\mathcal{N})$ are mutually
independent random variables, $\alpha_i \in
\mathcal{A}=(0,A_{max}),\,A_{max}\in R^{+}$, drawn from the same
distribution function $F(\cdot)$ with finite density $f(\cdot)$ on
$\mathcal{A}$. We assume that it holds: $U_{1}(\alpha_i)\geq
U_{2}(\alpha_i)\geq \ldots \geq U_{K}(\alpha_i)\geq0$, for each
$\alpha_i \in \mathcal{A}$, $i\in\mathcal{N}$. The SO $i$ pays for
the assigned channels an amount of money that is determined by the
PO.

\textbf{Primary Operators:} Each $j^{th}$ PO receives $K_{cj}$
channels from the CO at a cost of $Q(K_{cj})$ monetary units and
decides how many he will reserve for his own needs, $K_{j0}$, and
how many he will allocate to each one of the $N$ SOs at his
secondary market, $\mathbf{K}_j=(K_{ji}:\,i\in\mathcal{N})$. We
assume that the valuation of the $j^{th}$ PO for acquiring the
$k^{th}$ channel is $V_{k}(p_j)\,\in\,\mathcal{R}^{+}$ which belongs
to a known family of functions $V_{k}(\cdot)$ and is parameterized
by the private variable $p \in \mathcal{P}=(0,P_{max}),\,P_{max}\in
R^{+}$. In analogy with $\alpha$, $p$ is the type of the PO and
models his spectrum needs. The valuation functions are considered
positive, monotonically increasing and continuously differentiable
w.r.t. the type $p_j$ and we assume that it is: $V_{1}(p_j)\geq
\ldots \geq V_{K}(p_j)\geq0$. The benefit of the PO from reselling
his spectrum is given by the revenue component $H(\mathbf{K}_j)$
which depends on the number of leased channels. We define the
combined valuation - revenue objective of each PO $j\in\mathcal{M}$
as follows:
\begin{equation}
\hat{V}(p_j,K_{j0},\mathbf{K}_j)=\sum_{k=1}^{K_{j0}}V_{k}(p_j)+H(\mathbf{K}_j)
\label{eq:PO_combined_objective}
\end{equation}

\textbf{Controller:} The goal of the controller is to increase the
spectrum utilization while ensuring the viability of the secondary
markets. Therefore he acts as regulator and deploys an incentive
mechanism to induce a channel allocation that maximizes a balanced
sum of the POs' combined objectives and the valuations of the SOs:
\begin{equation}
C(\beta)=\sum_{j=1}^{M}[\hat{V}(p_j,K_{j0},\mathbf{K}_j)+\beta\sum_{i=1}^{N}\sum_{k=1}^{K_{ji}}U_k(\alpha_i)]
\label{eq:CO_balanced objective}
\end{equation}
where $\beta\in R^{+}$ is defined by the CO and determines this
balance. Apparently, as $\beta$ increases, the allocation of
spectrum will favor the SOs. Notice that the objective of the CO
incorporates both the channel valuation of the POs and their revenue
components, since the latter are the their motivation for
reallocating the spectrum.


\section{Unregulated Hierarchical Spectrum Allocation}
\label{section:unregulated}

We begin our study with the unregulated hierarchical spectrum
allocation. In this case, the CO does not take into account the
secondary demand when he allocates the channels to the POs. The
latter, may also be oblivious to the secondary demand at the moment
they ask for spectrum, or they can make an early conjecture for the
SOs needs or even they can be aware of the exact secondary demand.
For all these scenarios, we show that the unregulated spectrum
allocation induces efficiency loss. The model and analysis of this
section is used in the sequel to introduce our mechanism and assess
its efficacy.

\subsection{First Stage: POs - CO Interaction}
In the first stage, the POs ask for spectrum and the controller
determines the channel distribution, $\mathbf{K}_c=(K_{cj}:\,
j\in\mathcal{M})$ and the payment $Q(K_{cj})$. We assume that the CO
knows the family of the valuation functions of the POs,
$V_k(p),\,k\in\mathcal{K}$ but not their exact types,
\cite{krishna}, \cite{nisan}. Therefore the CO runs a
Vickrey-Clarke-Groove (VCG) auction, which is the most prominent
efficient auction \cite{krishna}, in order to elicit this
information. Every PO $j\in\mathcal{M}$ submits a scalar bid,
$r_j\in \mathcal{P}$, in order to declare his type. The CO collects
these bids, $\mathbf{r}=(r_j:\,j\in\mathcal{M})$, and finds the
channel allocation that maximizes the total valuation of all the POs
by solving the \textbf{CO Spectrum Allocation Problem,}
($\mathbf{P_{co}}$):
\begin{equation}
\max_{\mathbf{K}_c}\sum_{j=1}^{M}\sum_{k=1}^{K_{cj}} V_k(r_j)
\end{equation}
s.t.
\begin{equation}
\sum_{j=1}^{M}K_{cj}\leq K,\,\,K_{cj}\in\{0,1,2,\ldots,K\}
\end{equation}
One simple method to find the solution, $\mathbf{K}_{c}^{*}$, of
problem $\mathbf{P_{co}}$, is to sort the valuations of the POs
$V_k(r_j)$, $j\in\mathcal{M}$, $k\in\mathcal{K}$, in decreasing
order and allocate the channels to the primary operators with the
$K$ highest valuations. Apparently, the number of channels the
$j^{th}$ PO receives depends both on his own bid $r_j$ and the bids
of the other POs, $r_{-j}=(r_m:\,m\in\mathcal{M}\setminus\{j\})$,
i.e. $K_{cj}(r_j,\mathbf{r}_{-j})$.

The payment imposed to each PO, according to the VCG payment rule
\cite{krishna}, is equal to the externality he creates to the other
POs:
\begin{equation}
Q(r_j,r_{-j})=\sum_{m\neq
j}^{M}\sum_{k=1}^{\tilde{K}_{cm}^{*}}V_k(r_m) - \sum_{m\neq
j}^{M}\sum_{k=1}^{K_{cm}^{*}} V_k(r_m)\label{eq:PO_payment}
\end{equation}
where $K_{cm}^{*}$ is the number of channels allocated to each PO
$m\in\mathcal{M}\setminus\{j\}$ according to the solution of problem
($\mathbf{P_{co}}$) and $\tilde{K}_{cm}^{*}$ the respective number
when the $j^{th}$ PO does not participate in the auction, i.e.
$r_j=0$.

Let us assume now that the POs bid without knowing the secondary
demand. In this case, they consider only the benefit from using the
acquired channels for their own needs, $K_{j0}=K_{cj}$, and hence
they determine their bid by solving the \textbf{PO Bidding Problem},
($\mathbf{P_{po}^{b}}):$
\begin{equation}
r_{j}^{*}=\arg\max_{r_{j}}\{\sum_{k=1}^{K_{cj}(r_j,r_{-j})}
V_k(p_{j})-Q(r_j,r_{-j})\} \label{eq:PO_bidding}
\end{equation}
Since VCG auctions are incentive compatible, \cite{krishna}, the POs
will reveal their actual types, $r_{j}^{*}=p_j$, $\forall
j\in\mathcal{M}$. However, if POs are aware of the secondary demand
(or can make a conjecture) then they will bid so as to maximize
their combined valuation - revenue objective:
\begin{equation}
r_{j}^{*}=\arg\max_{r_{j}}\{\hat{V}(p_j,K_{j0},\mathbf{K}_j)
-Q(r_j,r_{-j})\} \label{eq:PO_bidding_2}
\end{equation}
Apparently, if the CO still uses the payment rule given by eq.
(\ref{eq:PO_payment}) then the auction in this stage is not truthful
anymore.

\subsection{Second Stage: SOs - PO Interaction} \label{sec:section3b}
In the second stage of the hierarchical spectrum allocation, each
$j\in\mathcal{M}$ PO finds the optimal allocation,
$(\mathbf{K}_{j}^{*},K_{j0}^{*})$, of his $K_{cj}$ channels that
maximizes his combined objective, eq.
(\ref{eq:PO_combined_objective}). This is given by the solution of
the \textbf{PO Spectrum Allocation Problem}, ($\mathbf{P_{po}}$):
\begin{equation}
\max_{\mathbf{K}_j,K_{j0}} \hat{V}(p_j,K_{j0},\mathbf{K}_j)
\label{eq:P_po1}
\end{equation}
s.t.
\begin{equation}
K_{j0}+\sum_{i=1}^{N}K_{ji}\leq
K_{cj},\,K_{ji},\,K_{j0}\in\{0,1,2,\ldots,K_{cj}\} \label{eq:P_po2}
\end{equation}

We assume that after receiving his spectrum from the CO, each PO
obtains only partial information about the underneath secondary
market. Namely, he learns the family of the SOs functions
$U_k(\alpha),\,k\in\mathcal{K}$, and the types distribution function
$F(\cdot)$ but not the actual SOs types. To elicit this missing
information the PO runs an \emph{optimal auction} where each one of
the $N$ SOs submits a bid, $b_i\in \mathcal{A}$ in order to declare
his type $\alpha_i$. The PO collects the bids,
$\mathbf{b}=(b_i:\,i\in\mathcal{N})$, and determines the allocated
spectrum and the respective payment for each bidder. Here, the
seller (PO) is also interested in the auctioned items and hence he
compares his possible revenue from selling a channel with the
valuation for using it, $V_{k}(\cdot)$ before he decides if he will
allocate it to a SO or reserve it, \cite{myerson},\cite{la}.

The maximization of the expected revenue, ($\mathbf{P_{po}}$), can
be transformed to a deterministic channel allocation problem. Let us
first define the additional expected revenue the PO incurs for
assigning the $k^{th}$ channel to the $i^{th}$ SO. In auction
theory, \cite{branco}, this is known as the \emph{contribution} of
the bidder and is defined as:
\begin{equation}
\pi_{k}(b_i)=U_k(b_i)-\frac{dU_k(\alpha)}{d\alpha}\vline_{\alpha=b_i}\frac{1-F(b_i)}{f(b_i)}
\end{equation}
where $F(\cdot)$ and $f(\cdot)$ are the cdf and pdf of the SOs. If
these contributions are monotonically strictly increasing in the SOs
types and decreasing in the number of channels, then they satisfy
the so-called \emph{regularity conditions}, \cite{branco}, and the
auction problem $\mathbf{P_{po}}$ is called \emph{regular}. In this
case the channel allocation that maximizes the combined objective of
the $j^{th}$ PO can be easily derived using the following
deterministic allocation and payment rules.

\subsubsection{PO Optimal Auction Allocation Rule} The auctioneer (PO
$j$) calculates the contributions $\pi_k(b_i)$ of each SO
$i\in\mathcal{N}$ for all the auctioned channels,
$k=1,\ldots,K_{cj}$, and selects the $K_{cj}$ highest of them. In
the sequel, he compares these $K_{cj}$ contributions with his own
valuations for the channels and constructs the
contribution-valuation vector $X_j$ which has $K_{cj}$ elements in
decreasing order:
\begin{equation}
X_j=(x_{(l)}:\,x_{(l)}>x_{(l+1)},\,l=1,\ldots,K_{cj})
\end{equation}
Then, the PO simply assigns each channel $l=1,\ldots,K_{cj}$ to the
respective $i^{th}$ SO if $x_{(l)}=\pi_k(b_i)$ or he reserves it for
himself if $x_{(l)}=V_{k}(p_j)$. For example, for a PO with $4$
channels and two SOs bidders, a possible instance of $X_j$ is
$X_j=(V_1(p_j),\pi_1(b_1),\pi_1(b_2),V_2(p_j))$ which means that the
PO will reserve $2$ channels for himself and assign one to each SO.

\subsubsection{PO Optimal Auction Payment Rule} The price that each
SO $i$ pays for receiving the $k^{th}$ spectrum channel depends on
the bids submitted by all the other SOs,
$b_{-i}=(b_n:\,n\in\mathcal{N}\setminus\{i\})$. Namely, let us
denote with $z_{k}(b_{-i})$ the minimum bid that the $i^{th}$ SO has
to submit in order to acquire the $k^{th}$ channel, \cite{branco}:
\begin{equation}
z_{k}(b_{-i})=\inf \{ \hat{\alpha}_i \in \mathcal{A} :
\pi_k(\hat{\alpha}_i) \geq \max \{0, x_{(K_{cj}+1)}\} \}
\end{equation}
This means that in order to get the $k^{th}$ item the $i^{th}$ SO
has simply to submit a bid high enough to draft his contribution
within the first $K_{cj}$ elements of $X_j$. The actual charged
price for each channel is equal to his valuation had he a type equal
to this minimum bid. Hence the aggregate payment for the SO is:
\begin{equation}
h(b_i,b_{-i})=\sum_{k=1}^{K_{ji}(b_i,b_{-i})}U_{k}(z_{k}(b_{-i}))\label{eq:SO_payment}
\end{equation}

This payment rule is an extension of the original rule introduced in
\cite{myerson} and \cite{branco} and has been used also for the case
that the seller has valuation for the auctioned items in \cite{la}.
Hence, each SO $i\in\mathcal{N}$ bids according to the \textbf{SO
Bidding Problem}, ($\mathbf{P_{so}}$):
\begin{equation}
b_{i}^{*}=\arg\max_{b_i}\{\sum_{k=1}^{K_{ji}(b_i,b_{-i})}U_k(\alpha_i)-h(b_{-i})\}
\label{eq:SO_bidding}
\end{equation}
Due to the payment and the respective monotonic allocation rule, the
auction mechanism is incentive compatible and individual rational,
\cite{jia}, \cite{nisan}, hence $b_{i}^{*}=\alpha_i$, $\forall
i\in\mathcal{N}$.

\subsection{Inefficiency of the Unregulated Hierarchical Scheme}
\label{sec:section3c}

The first reason that renders inefficient this hierarchical
allocation is the \textbf{coordination problem}: the CO allocates
the channels to the POs by solving $\mathbf{P_{co}}$ without
considering the demand of the SOs. In case the POs are also unaware
of the secondary demand the auction in the first stage is truthful
and efficient wrt the POs needs but not wrt to the POs - SOs  joint
spectrum requirements. Namely, the CO may allocate more channels to
a PO who is going to encounter smaller secondary demand than another
PO. Now assume that the POs receive the bids of the SOs or make an
early conjecture about the secondary demand before they ask CO for
spectrum. In this case, if the CO is still unaware of the SOs needs,
the hierarchical spectrum allocation is even more inefficient since
the POs will bid in order to maximize their combined objective,
according to eq. (\ref{eq:PO_bidding_2}), and not their valuations,
eq. (\ref{eq:PO_bidding}). This means that in the first auction, the
seller (CO) and the bidders (POs) will use different valuation
functions for determining the prices and the bids respectively.
Therefore, the auction will not be incentive compatible anymore,
i.e. $r_j\neq p_j$, $j\in\mathcal{M}$.

At the same time, and independently of the coordination problem,
there exists the \textbf{objectives misalignment problem}. The CO
and the POs have different goals and the revenue maximizing auction
that is organized by the latter incurs efficiency loss,
\cite{krishna}. In the setting we study, this means that a PO may
reserve a channel for himself while there exist SOs with higher
valuation for it. Finally, notice that even if both the CO and the
POs are aware of the exact secondary demand, still the problem of
their objectives misalignment exists and renders inefficient the
hierarchical spectrum allocation.

\section{Regulated Hierarchical Spectrum
Allocation}\label{section:regulatedspectrum}

In this section we build upon the previous analysis and introduce
our incentive mechanism. First, we explain the basic idea of the
mechanism and the difficulties that the controller encounters in
applying it. Next we introduce the $\beta$-optimal auction which is
required in order to enable the POs to balance their revenue and the
efficiency of their spectrum redistribution. Finally, we discuss the
efficacy of the mechanism and its requirements.

\subsection{Incentive Mechanism $\mathcal{M}_{R}$}
The goal of the controller is to induce the channel allocation
$\{K_{j0}^{*},\mathbf{K}_{j}^{*}\}$ for each PO $j\in\mathcal{M}$
and the respective secondary market that maximizes his objective
$C(\beta)$, given by eq. (\ref{eq:CO_balanced objective}). This
allocation stems from the solution of the \textbf{CO Balanced
Spectrum Allocation Problem,} ($\mathbf{P_{co}^{bal}}$):
\begin{equation}
\max_{\{\mathbf{K}_{j0},\mathbf{K}_j\}}\sum_{j=1}^{M}\left[\hat{V}(p_j,K_{j0},\mathbf{K}_j)+\beta\sum_{i=1}^{N}\sum_{k=1}^{K_{ji}}U_k(\alpha_i)\right]
\label{eq:CO_balanced_objective_problem}
\end{equation}
s.t.
\begin{equation}
\sum_{j=1}^{M}[K_{j0}+\sum_{i=1}^{N}K_{ji}]\leq K_{c},\, K_{j0},
K_{ji}
\in\{0,1,2,\ldots,K_c\}\label{eq:CO_balanced_objective_problem2}
\end{equation}
parameter $\beta\in R^{+}$ is determined by the CO and defines
implicitly the revenue of the POs and the welfare of the SOs.

The difficulties the controller encounters to achieve his goal, are
three: \textbf{(i)} the CO is not aware of the types of the POs,
$p_j,\,j\in\mathcal{M}$, \textbf{(ii)} he does not know the types of
the SOs in each secondary market, $a_i,\,i\in\mathcal{N}$, and
\textbf{(iii)} he cannot directly dictate the POs how to
redistribute their channels. In economic terms, conditions (i) and
(ii) capture the \emph{hidden information asymmetry},
\cite{courcoubetis}, of the spectrum market which means that the
controller is not aware of the actual needs of the operators.
Similarly, condition (iii) describes the \emph{hidden action
asymmetry}, which exists in the market because the CO is not aware
of the actions of the POs. The introduced incentive mechanism, which
we call \emph{Mechanism} $\mathcal{M_R}$, eliminates these
asymmetries and achieves the desirable spectrum allocation.

The proposed scheme is based on pricing and the underlying idea is
that the controller creates a coupling between the spectrum
allocation decisions of the POs and their cost for acquiring the
spectrum in order to bias their revenue maximizing strategy. Namely,
we assume that the CO reimburses the $j^{th}$ PO with the following
price:
\begin{equation}
L_j(\mathbf{\alpha},\mathbf{K}_j,\beta)=\beta\sum_{i=1}^{N}\sum_{k=1}^{K_{ji}}U_k(\alpha_{i})
\label{eq:regulation_fee}
\end{equation}
This modifies the PO's objective function as follows:
\begin{equation}
\hat{V}_{R}(p_j,K_{j0},\mathbf{K}_j,\beta)=\hat{V}(p_j,K_{j0},\mathbf{K}_j)+\beta\sum_{i=1}^{N}\sum_{k=1}^{K_{ji}}U_k(\alpha_{i})
\label{eq:PO_bal_objective}
\end{equation}
$\hat{V}_{R}(\cdot)$ is the regulated new combined objective of each
PO which depends on parameter $\beta$ and is aligned with the
balanced objective of the CO, eq. (\ref{eq:CO_balanced objective}).

\subsection{The $\beta$-Optimal Auction Mechanism}
Each PO maximizes $\hat{V}_{R}(\cdot)$ by solving a new allocation
problem $\mathbf{P_{po}^{\beta}}$ which differs from the respective
$\mathbf{P_{po}}$ problem in the objective function that is given
now by eq. (\ref{eq:PO_bal_objective}). Since the types of the SOs
are unknown, the primary operator runs again an auction to elicit
this hidden information. Nevertheless, this is neither an efficient
nor an optimal auction and hence he cannot employ any of the known
auction schemes. To address this problem, we introduce a new
multi-item auction mechanism, the \textbf{$\beta$-optimal} auction,
which ensures the maximization of the balanced objective defined in
eq. (\ref{eq:PO_bal_objective}). This mechanism is similar to the
optimal auction discussed in section \ref{sec:section3b} with the
difference that the allocation rule is biased by parameter $\beta$.
This modification affects the channels allocation and results in
reduced revenue for the auctioneer but improved allocation
efficiency. The combination of optimal and efficient auctions has
been also suggested in \cite{sandhold} for single item allocation
where the authors proposed an efficient auction with a lower bound
on the seller's revenue.


Let us now explain the rationale and machinery of the
$\beta$-optimal auction. First we define the $\beta$-contribution
for each SO $i\in\mathcal{N}$ under a certain PO $j\in\mathcal{M}$,
as follows:
\begin{equation}
\pi_{k}^{\beta}(b_i)=(1+\beta)U_k(b_i)-\frac{dU_k(\alpha)}{d\alpha}\vline_{\alpha=b_i}\frac{1-F(b_i)}{f(b_i)}
\end{equation}
Since $\beta\geq 0$ it will be
$\pi_{k}^{\beta}(\alpha_i)\geq\pi_{k}(\alpha_i)$ for all the SOs and
all the channels. Moreover, notice that if the initial contributions
satisfy the \emph{regularity} conditions, \cite{branco}, then the
$\beta$-contributions will also satisfy them and hence problem
$\mathbf{P_{po}^{\beta}}$ will be regular. Therefore, we can again
derive deterministic channel allocation and payment rules.

\textbf{$\beta$-Optimal Auction Allocation Rule:} Similarly to the
allocation rule of the optimal auction, the $j^{th}$ PO calculates
the $\pi_{k}^{\beta}(b_i)$ for all the SOs $i\in\mathcal{N}$ and all
the channels $k=1,\ldots,K_{cj}$ and compares them with his own
valuations in order to construct the contribution-valuation vector
$X_{j}^{\beta}$:
\begin{equation}
X_{j}^{\beta}=(x_{(l)}^{\beta}:\,x_{(l)}^{\beta}>x_{(l+1)}^{\beta},\,l=1,\ldots,K_{cj}^{\beta})
\end{equation}
Using $X_{j}^{\beta}$ the PO allocates his channels to the
respective $K_{cj}$ highest contributions and valuations. The
resulting channel allocation
$(K_{j0}^{\beta},\mathbf{K}_{j}^{\beta})$ solves problem
$\mathbf{P_{po}^{\beta}}$ and maximizes the new regulated objective
$\hat{V}_{R}(p_j,K_{j0}^{\beta},\mathbf{K}_{j}^{\beta},\beta)$.
Again, this allocation rule is monotone increasing in the types of
the SOs.

\textbf{$\beta$-Optimal Auction Payment Rule:} The payment rule
changes in order to comply with the new allocation rule. Namely, the
minimum bid that the $i^{th}$ SO needs to submit in order to acquire
the $k^{th}$ channel is:
\begin{equation}
z_{k}^{\beta}(b_{-i})=\inf \{ \hat{\alpha}_i \in \mathcal{A} :
\pi_{k}^{\beta}(\hat{\alpha}_i) \geq \max \{0,
x_{(K_{cj}^{\beta}+1)}^{\beta}\} \}
\end{equation}
and, similarly to the previous mechanism, the total payment for this
SO is :
\begin{equation}
h^{\beta}(b_i,b_{-i})=\sum_{k=1}^{K_{ji}^{\beta}(b_i,b_{-i})}U_{k}(z_{k}^{\beta}(b_{-i}))\label{eq:SO__beta_payment}
\end{equation}
Under this new auction mechanism, each SO $i\in\mathcal{N}$ selects
his bid so as to maximize his payoff, (\textbf{SO $\beta$-Bidding
Problem,} $\mathbf{P_{so}^{\beta}}$):
\begin{equation}
b_{i}^{*}=\arg\max_{b_i}\{\sum_{k=1}^{K_{ji}^{\beta}(b_i,b_{-i})}U_k(\alpha_i)-h^{\beta}(b_{-i})\}
\label{eq:SO_bidding}
\end{equation}

This new auction mechanism improves the efficiency of the POs - SOs
interaction and at the same time preserves the necessary properties
of the classical optimal auctions.
\begin{mydef}
The $\beta$-optimal auction mechanism preserves the incentive
compatibility and the individual rationality properties of the
optimal multi-unit auction introduced in \cite{branco}.
\end{mydef}
The proof of this proposition can be found in our technical report,
\cite{iosifidis-report}.


\begin{figure}[t]
\begin{center}
\epsfig{figure=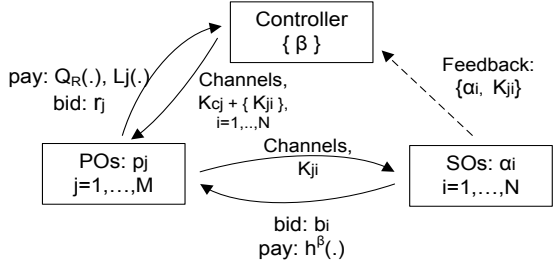, width=7.6cm,height=3.8cm}
\end{center}
\caption{The machinery of incentive mechanism $\mathcal{M}_R$. The
circulated information, bids and channel allocation among the SOs,
POs and the Controller is depicted. The feedback can be provided
from the SOs to the CO directly, or be inferred using other means.}
\label{fig:explain}
\end{figure}

\begin{algorithm}[t]
\caption{(Mechanism $\mathcal{M}_{R}$)}%
\label{alg1}
\begin{algorithmic}
\STATE \textbf{1st Stage:} \emph{Channel Allocation by the CO ($\beta$ is given)}.\\%
\STATE(\textbf{1.1:}) Each SO $i$ bids to the respective PO,
according to problem $\mathbf{P_{so}^{\beta}}$, eq.
(\ref{eq:SO_bidding}) and informs the CO about his needs (\emph{feedback} $\alpha_i$).\\%
\STATE(\textbf{1.2:}) Each PO bids to the CO, by solving $\mathbf{P_{po}^{b_{\beta}}}$, eq. (\ref{eq:PO_bal_bidding}).\\%
\STATE(\textbf{1.3:}) The CO solves problem $\mathbf{P_{co}^{bal}}$,
eq.
(\ref{eq:CO_balanced_objective_problem})-(\ref{eq:CO_balanced_objective_problem2}),
and allocates
$K_{cj}^{\beta}=K_{j0}^{\beta}+\sum_{i=1}^{N}K_{ji}^{\beta}$
channels to each PO $j\in\mathcal{M}$.\\%
\STATE \textbf{2nd Stage:} \emph{Channel Redistribution by the POs}.\\%
\STATE(\textbf{2.1:}) Each PO $j\in\mathcal{M}$ redistributes his
channels according to the
\emph{$\beta$-Optimal Allocation Rule}, $X_{j}^{\beta}$.\\%
\STATE (\textbf{2.2:}) Each SO $i$ reveals to the
CO the allocation decisions of the respective PO (\emph{feedback} $K_{ji}^{\beta}$).\\%
\STATE (\textbf{2.3:}) The CO collects the feedback, calculates the
price $L_j(\cdot)$, eq. (\ref{eq:regulation_fee}), and determines
the overall payment for each PO:
\begin{equation}
\Lambda_{j}=Y-\beta[\sum_{i=1}^{N}\sum_{k=1}^{K_{ji}^{\beta}}U_k(\alpha_{i})]+
Q_{R}(r_j,\mathbf{r}_{-j}) \nonumber
\end{equation}
Parameter $Y\in\mathcal{R}^{+}$ is a properly selected offset in
order to render the payments positive.
\end{algorithmic}
\end{algorithm}

\subsection{Efficacy and Requirements of Mechanism $\mathcal{M}_{R}$}

The pricing that is imposed by the CO, eq.
(\ref{eq:regulation_fee}), not only bias the channel distribution
strategy of the POs, but also changes their bidding strategy.
Namely, each PO $j\in\mathcal{M}$ after receiving the bids of his
SOs, determines his optimal bid by solving the \textbf{PO
$\beta$-Bidding Problem,} ($\mathbf{P_{po}^{b_{\beta}}}$):
\begin{equation}
r_{j}^{*}=\arg\max\{\hat{V}_{R}(p_j,K_{j0}^{\beta},\mathbf{K}_{j}^{\beta})-Q_{R}(r_j,r_{-j})
\} \label{eq:PO_bal_bidding}
\end{equation}
The CO determines the channel allocation by solving problem
$\mathbf{P_{co}^{bal}}$, eq.
(\ref{eq:CO_balanced_objective_problem}) -
(\ref{eq:CO_balanced_objective_problem2}), and calculates the new
VCG prices as follows:
\begin{equation}
Q_{R}=\sum_{m\neq
j}^{M}\hat{V}_{R}(r_m,\tilde{K}_{m0}^{\beta},\tilde{\mathbf{K}}_{m}^{\beta})
- \sum_{m\neq
j}^{M}\hat{V}_{R}(r_m,K_{m0}^{\beta},\mathbf{K}_{m}^{\beta})
\label{eq:PO_payment_R1}
\end{equation}
Again, the number of channels allocated to each PO $m$,
$(K_{m0}^{\beta}, \mathbf{K}_{m}^{\beta})$, depends on bids
submitted by all the POs. Also, $(\tilde{K}_{m0}^{\beta},
\tilde{\mathbf{K}}_{m}^{\beta})$ is the channel allocation when
$r_j=0$. Therefore, the POs are induced to bid truthfully,
$r_{j}^{*}=p_j,\,j\in\mathcal{M}$. Apparently, $\mathcal{M_R}$
solves both the coordination and the misaligned objectives problem.
The improved efficiency of the channel allocation under the
$\beta$-optimal auction, can be also realized by considering the
inequality $U_k(\alpha_i)\geq\pi_{k}^{\beta}(b_i)\geq \pi_{k}(b_i)$
which holds for all the POs and SOs. We summarize mechanism
$\mathcal{M}_{R}$ in Algorithm $\mathbf{1}$ and Figure
\ref{fig:explain}.

In order to calculate the prices $L_j(\cdot)$, the CO needs to know
the actual types of the SOs in the respective secondary market and
the amount of spectrum that is allocated to them by the PO. There
are many different methods and scenarios about how the CO can
acquire this information. First, the SOs may directly provide it
through a feedback loop, Figure \ref{fig:ssa1}, if there is a
trusted relationship among them. Equivalently, the CO may be able to
observe the interaction (bidding) of the SOs with the respective PO.
Recall that the SOs bid truthfully due to the incentive
compatibility property of the $\beta$-optimal auction. Finally, the
POs may also reveal the outcome of their interactions with the SOs.

Since the controller is on top of this hierarchy and manages the
spectrum, we can easily consider many similar methods that will
allow him to receive direct or indirect feedback about the SOs - POs
interaction. However, in all these cases, the basic assumption of
our mechanism is that the SOs do not anticipate the impact of their
bids to the CO decisions (\emph{price takers}). That is, their
bidding strategy is not affected by the monitoring/observation by
the controller, which is rather expected due to the large number of
the SOs, i.e. $N\times M$. Finally, we assume that the SOs bid to
the POs before the latter ask for spectrum. If we relax this
assumption, the coordination problem is by default unsolvable, but
our mechanism still improves the hierarchical spectrum allocation by
addressing the objectives misalignment problem. This issue is
discussed in the next section, in the context of the dynamic
spectrum markets, where it is more prevalent.

\section{Regulation in Dynamic Spectrum
Markets}\label{section:discussionanddynamicmarkets}

Until now, we ignored the dynamic aspect of the problem in order to
facilitate the analysis and we focused on the novel balanced auction
scheme. That is, we implicitly assumed that the interaction of the
CO with the POs, and the interactions of the latter with the SOs are
performed in the same time scale. This is a realistic assumption
since the current suggestions about the spectrum policy reform
advocate a more fine grained spatio-temporal management by the
regulators, \cite{peha2}. Nevertheless, the proposed mechanism
$\mathcal{M}_{R}$ can be extended for the case where the CO-POs and
POs-SOs interactions are realized in different time scales. Due to
lack of space we will briefly explain how the mechanism is adapted
for this setting.

Assume that the time is slotted and divided in time periods,
$\mathcal{I}=1,2,\ldots$, where each period is further divided in
$T$ time slots, $t=1,2,\ldots,T$. The CO determines his $K_c$
channels allocation in the beginning of each period while the POs
redistribute them in every slot. The CO $\mathbf{P_{co}^{bal}}$
problem for this setting is related to the spectrum allocation for
all the $T$ slots within each period:
\begin{equation}
\max_{\{K_{j0}^{t},\mathbf{K}_{j}^{t}\}}\sum_{t=1}^{T}\sum_{j=1}^{M}\left[\hat{V}(p_j,K_{j0}^{t},\mathbf{K}_{j}^{t})+\beta\sum_{i=1}^{N}\sum_{k=1}^{K_{ji}^{t}}U_k(\alpha_{i}^{t})\right]
\end{equation}
s.t.
\begin{equation}
\sum_{j=1}^{M}[K_{j0}^{t}+\sum_{i=1}^{N}K_{ji}^{t}]\leq K_{c},\,
t=1,\ldots,T
\end{equation}
where we have marked with the superscript $t$ the variables that
change in each slot. Obviously the CO cannot allocate the spectrum
optimally to the POs for the entire period since he is not aware of
the future demands of the SOs. Additionally, even if the CO had this
information, he could not determine the allocated channels to the
POs, $K_{cj}$, once in each period since these should be adapted to
the dynamic secondary demand, $K_{ji}^{t}$. Apparently, the
\textbf{coordination problem} cannot be solved optimally in this
setting.

Nevertheless, the CO is still able to solve the \textbf{objectives
misalignment problem} and induce the POs to allocate their spectrum
more efficiently. Assume that the CO-POs interaction is accomplished
either without taking into account the secondary demand as in
Section \ref{section:unregulated} or by considering the average
demand of the SOs, $\bar{\alpha}_i$. This will result in a certain
suboptimal channel allocation
$\bar{K}_{c}=\{\bar{K}_{cj}:\,j\in\mathcal{M}\}$. Then, in each slot
as the SOs demand will be realized, they will bid to the POs and at
the same time the CO will receive feedback (directly or indirectly)
about their needs. This way, the controller will be able to
determine the prices $L_j(\cdot)$ for each PO $j\in\mathcal{M}$ at
the end of the entire time period:
\begin{equation}
L_{j}^{T}(\mathbf{\alpha},\{\mathbf{K}_{j}^{t}\},\beta)=\sum_{t=1}^{T}\beta\sum_{i=1}^{N}\sum_{k=1}^{K_{ji}^{t}}U_k(\alpha_{i}^{t})
\end{equation}
Obviously, this subsequent pricing at the end of each time period
will induce the POs to allocate their spectrum by solving problem
$\mathbf{P_{po}^{\beta}}$ and maximizing eq.
(\ref{eq:PO_bal_objective}), instead of problem $\mathbf{P_{po}}$,
eq. (\ref{eq:P_po1})-(\ref{eq:P_po2}), in each time slot. Therefore,
the efficiency loss will be reduced.

\section{Simulation Results}\label{section:simulations}
In order to obtain insights about the proposed mechanism
$\mathcal{M}_R$, we simulate a representative three-layer
hierarchical market with one CO, $M=2$ POs and $N=10$ SOs under each
PO. We assume that the POs valuation functions for the $k^{th}$
channel are $V_k(p_j)=p_j/k$, where the types $p_j$ are uniformly
distributed in the interval $[5,6]$. Similarly, the SOs valuations
are $U_k(\alpha_i)=0.1\alpha_i/k$, and their types follow a uniform
distribution $F(x)=x/4$ on the interval $(0,4]$. The SOs
contributions are $\pi_k(\alpha_i)=(0.2\alpha_i-0.4)/k$ and the
respective $\beta$-contributions are
$\pi_{k}^{\beta}(\alpha_i)=[(0.2+\beta)\alpha_i-0.4]/k$. For each
random realization of the SOs and POs types, the results are
averaged over $40$ runs in order to capture the variance on the
spectrum demand.

\begin{figure}[htb]
\begin{center} \epsfig{figure=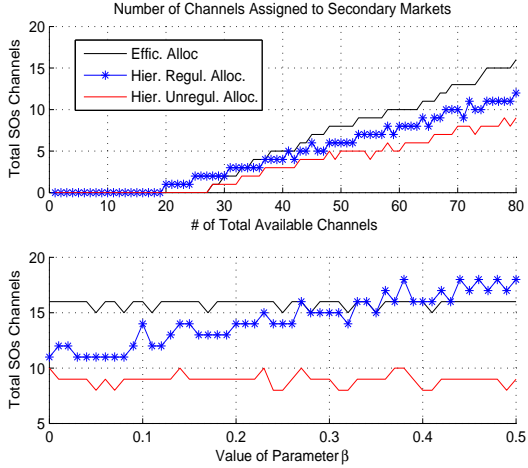,
width=8cm,height=6.5cm}
\end{center}
\caption{\emph{Upper Plot}: For $\beta=0.1$ the regulation mechanism
$\mathcal{M}_R$ increases the number of channels assigned to the
SOs. \emph{Lower Plot}: The SOs receive more channels for larger
values of $\beta$ ($K_c=80$).} \label{fig:eflos1}
\end{figure}

For our study we use as a benchmark the \emph{efficient} channel
allocation to the SOs. This allocation corresponds to the
hypothetical scenario where the CO would be able to assign directly
the channels to both the POs and the SOs and maximize the aggregate
spectrum valuations. In the upper plot of Figure \ref{fig:eflos1} we
show that in hierarchical unregulated market the number of total
channels assigned to the SOs is less than the channels in the
\emph{efficient} allocation. Mechanism $\mathcal{M}_R$ with
$\beta=0.1$ reduces this difference and increases the SOs channels.
Notice that the number of SOs channels is stil less than in the
\emph{efficient} allocation scenario, since the goal of the CO is
the combined revenue-efficiency balanced allocation.

\begin{figure}[htb]
\begin{center}
\epsfig{figure=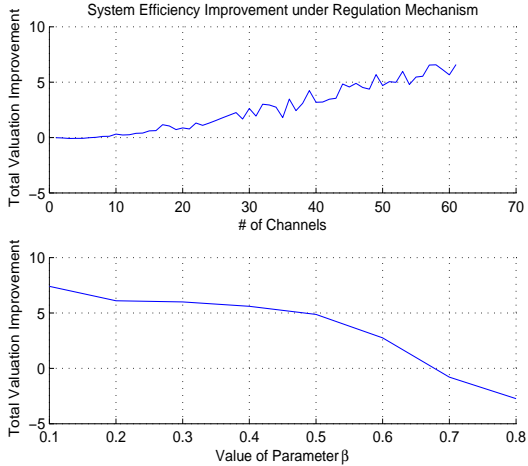, width=8cm,height=6.5cm}
\end{center}
\caption{\emph{Upper Plot}: The aggregate network efficiency (POs
and SOs valuations) increases with the $\mathcal{M}_R$, $\beta=0.1$,
$N=20$, $K_c=1:60$. \emph{Lower Plot}: For large values of $\beta$
the network efficiency decreases since the SOs are favored more than
the POs.} \label{fig:eflos2}
\end{figure}

In the same Figure we show that the number of channels assigned to
SOs vary with the value of $\beta$. Namely, when $\beta=0$ the
$\mathcal{M}_R$ regulated allocation is identical with the
unregulated allocation while for $\beta\approx0.35$ it reaches the
\emph{efficient} allocation. Notice that for larger values of
$\beta>0.37$ the SOs receive even more channels. This means that the
CO favors the SOs too much and render the channel allocation
inefficient. The impact of $\beta$ is depicted also in the lower
plot of Figure, \ref{fig:eflos2} where we see that for large values
the improvement in the aggregate valuation of the POs and SOs
becomes negative. For this plot, the number of SOs is $N=20$ and the
system welfare is maximized for $\beta=0.1$. If $\beta$ is further
increased, the welfare improvement decreases and eventually becomes
negative. Finally, we refer the interested readers to
\cite{iosifidis-report} for an additional, analytical example of
applying $\mathcal{M_R}$.

\section{Conclusions and Future Work}\label{section:conclusions}

In this paper we proposed an incentive mechanism that can be used by
the controller in order to regulate the interaction between the
primary and secondary operators in hierarchical spectrum markets and
to induce a new market equilibrium. This equilibrium depends on a
scalar parameter which is defined by the controller and determines
the efficiency of the secondary markets by adjusting the number of
channels allocated to the SOs. The mechanism is based on a novel
auction scheme which has a revenue-welfare balanced objective. There
exist several directions for future work. First, it is interesting
to relax the assumption of monopoly markets and study the impact of
POs competition on the network performance. We would like to compare
the results of competition and regulation in these hierarchical
markets. Another intriguing direction is to consider the more
realistic setting where there is no prior knowledge about the SOs
types or the family of POs and SOs valuation functions, and apply
learning schemes to elicit this hidden information in the spirit of
\cite{anil}. Finally, even more challenging is to consider the
scenario where the SOs anticipate the impact of their bidding to the
mechanism and strategize against it in order to gain higher
benefits.

\appendix

\subsection*{Numerical Example}

Consider a market where the CO has $12$ channels, there are $2$ POs
and $2$ SOs under each PO. The SOs types are drawn from a uniform
cdf $F(x)=x/2$ in the interval $(0,2]$ and their valuation for the
$k^{th}$ channel is $U_k(\alpha)=\frac{\alpha}{k}$. The respective
valuations of the POs are $V_k(p)=\frac{3*p}{k}$. We assume that
$p_1=1$ with $\alpha_1=1.2$ and $\alpha_2=1.5$ and $p_2=1.2$ with
$\alpha_3=1.3$ and $\alpha_4=1.4$. The contributions of the SOs are
$\pi_{k}(\alpha)=\frac{2\alpha-2}{k}$. If the channel allocation is
accomplished with the unregulated hierarchical method then in the
first stage the CO allocates the channels to the highest valuations
of the POs and these redistribute them comparing their own
valuations with the contributions of the SOs in their market. This
results in $Ch_{po}=10$ channels allocated to the POs and
$Ch_{so}=2$ assigned channels to the SOs. If however, the POs were
socially aware and considered the valuations of the SOs (instead of
their contributions) then the channel allocation would be $[Ch_{po},
Ch_{so}]=[8, 4]$. Finally, even this allocation is not the most
efficient because in the first stage the secondary demand has not be
considered. If for example the CO was able to allocate directly the
channels w.r.t. the POs and SOs valuations, then the allocation
would result in $[Ch_{po}, Ch_{so}]=[7, 5]$. Now, assume that we use
the the proposed mechanism $\mathcal{M}_{R}$, with $\beta=0.2$. In
this case, the number of assigned channels will be $Ch_{po}=9$ and
$Ch_{so}=3$, i.e. increased by $1$ for the SOs. Apparently for large
values of $\beta$ the allocation will favor the SOs and the revenue
of the POs will decrease.

\subsection*{Proof of Proposition $1$}

We focus on PO $j\in\mathcal{M}$ with $K_{cj}$ channels. We denote
$s_{ik}$ the probability of SO $i$ for receiving the $k^{th}$
channel which depends on the types of all the SOs. Additionally,
$c_i(\alpha_i)$ is the payment of each SO $i$ for all the channels
he acquired. \emph{Definition} $2$ and \emph{Lemma} $1$ in
\cite{branco} give the necessary conditions for the structure of the
bidders (SOs) valuation functions in order to ensure the (IC) and
(IR) properties. These conditions hold independently of the
objective of the auctioneer (PO) and hence they are not affected by
the incorporation of the linear term of the SOs valuation.

The objective of the PO w.r.t. the expected types of the SOs is:
\begin{align*}
E_{\mathcal{A}}[\hat{V}_{R}(\cdot)]&=\sum_{i=1}^{N}E_{\mathcal{A}}[c_i(\alpha_i)]+
\beta\sum_{i=1}^{N}E_{\mathcal{A}}[\sum_{k=1}^{K_{cj}}U_k(\alpha_i)s_{ik}]+\\
&+E_{\mathcal{A}}[\sum_{k=1}^{K_{cj}}V_k(p_j)(1-\sum_{i=1}^{N}s_{ik})]
\end{align*}
The first term is the payment by the SOs, the second is the pricing
and the third the valuation for the non-sold channels. After some
algebraic manipulations similarly to proof of \emph{Proposition} $1$
in \cite{branco}, we get:
\begin{align*}
&E_{\mathcal{A}}[\hat{V}_{R}(\cdot)]=\sum_{i=1}^{N}E_{\mathcal{A}}\{\sum_{k=1}^{K_{cj}}[(1+\beta)U_k(\alpha_i)-V_k(p_j)-
\frac{dU_k(\alpha_i)}{d\alpha}\\
&\frac{1-F(\alpha_i)}{f(\alpha_i)}]s_{ik}\}-\sum_{i=1}^{N}[\sum_{k=1}^{K_{cj}}U_k(0)-c_i(0)]+E_{\mathcal{A}}[\sum_{k=1}^{K_{cj}}V_k(p_j)]
\end{align*}
Using the necessary (IC) and (IR) conditions from \emph{Lemma} $1$
in \cite{branco}, it stems that the $\beta$-optimal payment rule is
given again by equation $(10)$ of \cite{branco}:
\begin{equation}
c_{i}^{*}(\alpha_i)=E_{\mathcal{A}}\{\sum_{k=1}^{K_{cj}}U_k(\alpha_i)s_{ik}-\int_{0}^{\alpha_i}\frac{dU_k(\alpha)}{d\alpha}s_{ik}d\alpha
\}
\end{equation}
where the probabilities of allocation are selected so as to maximize
the new objective of the auctioneer (instead of revenue only
maximization as in \cite{branco}). The optimal payment rule is the
one that yields zero payment and zero channel allocation for SOs
with zero type.

If the problem is \emph{regular} then the payment is as we described
in section \ref{section:regulatedspectrum} and the first term in the
PO's objective is maximized by using the $\beta$-optimal allocation
rule. This can be easily derived following the proof of the
respective \emph{Proposition} $2$ in \cite{branco}. Notice that if
the original respective problem in \cite{branco} is \emph{regular}
then also this modified problem is \emph{regular}. Apparently, the
inclusion of the SOs buyers valuations does not affect the
monotonicity of the allocation rule nor the critical value property
of the payment rule, \cite{nisan}. Hence, the modified auction is
still truthful.


\begin{thebibliography}{1}


\bibitem{zhao} Q. Zhao, and B. M. Sadler, ``A Survey of Dynamic Spectrum Access'',
\tit{IEEE Signal Proc. Mag.,} pp. 79-89, 2007.

\bibitem{peha2} J. M. Peha, ``Emerging Technology and Spectrum Policy Reform'',
\tit{Procf. of ITU Workshop on Market Mechanisms for Spectrum
Managememt,} 2007.

\bibitem{peha} J. M. Peha, and S. Panichpapiboon, ``Real-Time
Secondary Markets for Spectrum'', \tit{Telecommunications Policy,}
pp. 603-618, August 2004.

\bibitem{spectrumbridge} ``Spectrum bridge'',
\tit{http://www.spectrumbridge.com/.}

\bibitem{krishna} V. Krishna, \tit{Auction theory (2nd ed.),} Academic Press, 2010.

\bibitem{myerson} R. B. Myerson, ``Optimal Auction Design'',
\emph{Mathematics of Operations Research,} vol.6, pp.58-73, 1981.

\bibitem{hajek} V. Abhishek, and B. Hajek, ``Efficiency Loss in Revenue Optimal
Auctions'', \tit{IEEE Conf. Decision and Control, pp. 1082-1087,}
Dec. 2010.

\bibitem{maskin} E. S. Maksin, and J. G. Riley, ``Optimal Multi-unit Auctions'',
\emph{in Frank Hahn (ed) The Economics of Missing Markets,
Information and Games,} 1989.

\bibitem{branco} F. Branco, ``Multiple unit auctions of an indivisible good'',
\emph{Economic Theory,} vol.8, pp.77-101, 1996.

\bibitem{jia} J. Jia, Q. Zhang, Q. Zhang, and M. Liu, ``Revenue Generation for
Truthful Spectrum Auction in Dynamic Spectrum Access'', \tit{in
Proc. ACM Mobihoc,} 2009.

\bibitem{QZhang} L. Gao, X. Wang, Y. Xu, and Q. Zhang, ``Spectrum Trading in Cognitive Radio Networks: A
Contract-Theoretic Modeling Approach'', \tit{IEEE JSAC,} pp.
843-855, 2011.

\bibitem{zongpeng} A. Gopinathan, Z. Li, ``A Prior-Free Revenue
Maximizing Auction For Secondary Spectrum Access'', \tit{in Proc.
IEEE INFOCOM,} 2011


\bibitem{la} S.H. Chun, and R.J. La, ``Auction Mechanism for Spectrum Allocation
and Profit Sharing'', \tit{in Proc. of GameNets,} pp. 498-507, 2009.

\bibitem{courcoubetis} C. Courcoubetis, R. R. Weber, ``Pricing Communication
Networks: Economics, Technology and Modelling'', \tit{Wiley,} 2003.


\bibitem{niyato} D. Niyato, and E. Hossain, ``A Microeconomic Model for Hierarchical Bandwidth Sharing in
Dynamic Spectrum Access Networks: Distributed Implementation,
Stability Analysis, and Application'', \tit{IEEE Transactions on
Computers,} pp. 865-877, 2010.

\bibitem{huang} L. Duan, J. Huang, and B. Shou, ``Competition with Dynamic Spectrum Leasing'',
\tit{in Procc. of IEEE DySPAN,} 2010.

\bibitem{sarkar} G. Kasbekar, E. Altman, and S. Sarkar, ``A Hierarchical
Spatial Game over Licensed Resources'', \tit{in Proc. of GameNets,}
2009.

\bibitem{bitsaki} M. Bitsaki, G. D. Stamoulis, and C. Courcoubetis,
``An Efficient Auction-based Mechanism for Hierarchical Structured
Bandwidth Markets'', \tit{ Elsevier Computer Communications,} pp.
911-921, 2006.

\bibitem{nisan} N. Nisan, T. Roughgarden, E. Tardos, V. V. Vazirani,
``Algorithmic Game Theory'', \tit{Cambridge University Press,} 2007.

\bibitem{iosifidis-commag} G. Iosifidis, and I. Koutsopoulos, ``Challenges in Auction
Theory Driven Spectrum Management'', \tit{IEEE Communications
Magazine,} Vol. 49, No. 8 Aug. 2011.

\bibitem{sandhold} A. Likhodedov, and Tuomas Sandholm, ``Auction Mechanism for Optimally
Trading Off Revenue and Efficiency in Multi-unit Auctions'',
\emph{ACM Conference on EC,} 2004.

\bibitem{anil} A. K. Chorppath, and T. Alpcan, ``Learning User Preferences in Mechanism Design'',
\tit{to appear in IEEE CDC,} 2011.

\bibitem{iosifidis-report} G. Iosifidis, A. K. Chorppath, T. Alpcan, and G. Iosifidis, ``Incentive Mechanisms for
Hierarchical Spectrum Markets'', \tit{Technical Report, Arxiv...},
2011.

\end{thebibliography}
\end{document}